\pdfoutput=1
\documentclass[11pt]{article}
\usepackage[preprint]{acl}

\usepackage{times}
\usepackage{latexsym}
\usepackage{multirow}
\usepackage{graphicx}
\usepackage{amsmath}
\usepackage{amssymb}
\usepackage{booktabs}
\usepackage{subcaption}
\usepackage{caption}
\usepackage{bbding}
\usepackage{colortbl}
\usepackage{tcolorbox}

\newcommand{\ie}{\textit{i.e.}}
\newcommand{\eg}{\textit{e.g.}}
\usepackage{algorithm}
\usepackage{algorithmic}
\usepackage[T1]{fontenc}
\usepackage[utf8]{inputenc}

\usepackage{microtype}
\usepackage{bbold}

\usepackage{inconsolata}

\title{Sliding Windows Are Not the End: Exploring Full Ranking with Long-Context Large Language Models}
\author{Wenhan Liu$^1$, Xinyu Ma$^2$, Yutao Zhu$^1$, Ziliang Zhao$^1$, \textbf{Shuaiqiang Wang}$^2$ \\ 
\textbf{Dawei Yin}$^2$ \and \textbf{Zhicheng Dou}$^{1}$\thanks{Corresponding author.} \\
$^1$Gaoling School of Artificial Intelligence, Renmin University of China \\
$^2$Baidu Inc., Beijing, China \\
\texttt{lwh@ruc.edu.cn, xinyuma2016@gmail.com, dou@ruc.edu.cn}
}
\begin{document}
\maketitle

\begin{abstract}
Large Language Models (LLMs) have shown exciting performance in listwise passage ranking. 
Due to the limited input length, existing methods often adopt the \textit{sliding window} strategy.
Such a strategy, though effective, is inefficient as it involves repetitive and serialized processing, which usually re-evaluates relevant passages multiple times.
As a result, it incurs redundant API costs, which are proportional to the number of inference tokens.
The development of long-context LLMs enables the \textit{full ranking} of all passages within a single inference, avoiding redundant API costs.
In this paper, we conduct a comprehensive study of long-context LLMs for ranking tasks in terms of efficiency and effectiveness.
Surprisingly, our experiments reveal that full ranking with long-context LLMs can deliver superior performance in the supervised fine-tuning setting with a huge efficiency improvement.
Furthermore, we identify two limitations of fine-tuning the full ranking model based on existing methods: (1) sliding window strategy fails to produce a full ranking list as a training label, and (2) the language modeling loss cannot emphasize top-ranked passage IDs in the label.
To alleviate these issues, we propose a complete listwise label construction approach and a novel importance-aware learning objective for full ranking. 
Experiments show the superior performance of our method over baselines.
Our codes are available at \url{https://github.com/8421BCD/fullrank}.

% Furthermore, for the full ranking, we observe the incompleteness of the existing listwise label construction method, and the imbalance problem between positive and negative labels when outputting more labels.

% Long-context LLMs, which support longer input sequences, open up avenues for ranking all the passages in a single inference, which we call full ranking. They eliminate the need for sliding window strategy, thereby reducing redundant API costs. However, It remains unclear which of the two strategies is superior in terms of effectiveness and efficiency.
% In this paper, we explore for the first time the application of long-context LLMs for ranking tasks and conduct a series of experiments around these two strategies. Our empirical results reveal that the reranker based on full ranking strategy outperforms the reranker based on sliding window strategy in supervised fine-tuning scenario. We further propose an importance-aware loss that is better suited for fine-tuning ranking model compared to the traditional generative loss. Besides, based on existing long-context LLMs, our experiments prove that the full ranking strategy is significantly more efficient than the sliding window strategy.

% However, during the sliding window process, relevant passages can be inferenced multiple times as they bubble up from lower positions. This results in redundant API costs, which is proportional to the number of inference tokens. Besides, the serialization of this process also leads to an efficiency bottleneck. 

\end{abstract}

\section{Introduction}
\label{sec:intro}
% 大模型的优势，以及在ranking上应用效果很好。大概是咋做的，取得什么效果
In recent years, large language models (LLMs) have demonstrated impressive zero-shot capabilities in passage ranking tasks~\cite{rankgpt, llm4ir_survey}. The listwise ranking approach, which processes multiple passages simultaneously and directly outputs a reranked list of passage IDs, has been widely adopted and shown to outperform other methods. For instance, \citet{rankgpt} achieved state-of-the-art performance using the proprietary GPT-4 model on both the TREC~\cite{dl19} and BEIR~\cite{beir} benchmarks. Furthermore, several studies~\cite{rankvicuna, rankzephyr} also attempt to distill the listwise ranking capabilities of proprietary models into more moderately sized LLMs to enhance ranking efficiency and improve the reproducibility of results.

\begin{figure}[t]
	\centering
	\includegraphics[width=1\linewidth]{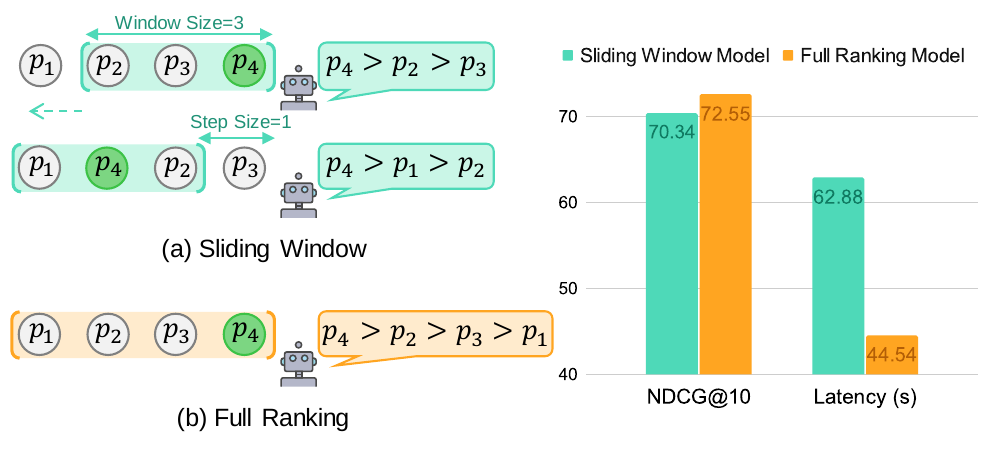}
	\caption{The sliding window strategy and full ranking strategy are shown in part (a) and part (b), respectively. The bar chart shows the comparison between our fine-tuned sliding window model and full ranking model in terms of NDCG@10 and latency (per query) on TREC DL19 dataset.}
	\label{fig:intro_example}
        % \vspace{-3mm}
\end{figure}

% 现在大模型for ranking的问题：主要是主流大模型input都比较短，导致只能输入a small group of passages，需要依赖一种滑动窗口策略。这种策略有重复计算等问题。
Due to the limited context length of many LLMs, existing listwise ranking methods can only process a subset of passages at a time, ultimately relying on sliding window strategy~\cite{rankgpt} to rank the entire passage list (see part (a) in Figure~\ref{fig:intro_example}). Based on fixed window size and step size, this strategy promotes relevant passages that are initially lowly ranked to the top. However, in this strategy, many passages are evaluated multiple times due to the overlapping part of adjacent windows, leading to significant redundancy and increased API costs which grow proportionally with the number of inference tokens. Additionally, the sequential dependency between windows also limits inference parallelization, resulting in efficiency bottlenecks.

% 随着长文本大模型的发展，full ranking 变成一种可能。full ranking具体怎么做，他在省钱上面有优势。但是他在效果和效率方面的效果不太clear
Recently, with the development of long-context techniques~\cite{Positional_Interpolation, LongNet}, some LLMs, such as Mistral-7B-Instruct-v0.3 (32k) and LLaMA 3.1-8B-Instruct (128k), have supported longer input lengths. This enables the input of all retrieved passages and the output of the full ranked list in a single step, which we refer to as \textit{full ranking} in this paper (see part (b) in Figure~\ref{fig:intro_example}). Full ranking not only eliminates the repetitive and time-consuming sliding window process but also reduces redundant passage inference, thereby significantly lowering API costs. Even though, the effectiveness and efficiency of the full ranking strategy remain unclear. As for effectiveness, although full ranking allows for more global interactions among passages which may lead to higher ranking accuracy, the increased input length also brings higher difficulty for LLMs. As for efficiency, while the full ranking eliminates the sliding window’s serial processing, the dramatically increased input length also increases the time cost of LLM encoding. We believe that a thorough investigation of these questions will facilitate the application of long-context LLMs to ranking tasks. 
% This is the primary focus of the work.

% \MXY{A comprehensive study of long-context LLMs for ranking}
% 讲讲在什么seeting下，做了啥探究。结论是什么。注意：突出我们的亮点
In this paper, we conduct a comprehensive study of long-context LLMs for ranking. Specifically, we compare the full ranking and sliding window strategies in terms of efficiency and effectiveness, and draw the following conclusions: (1) In zero-shot setting, the full ranking model demonstrates higher efficiency but lower effectiveness compared to the sliding window model. (2) In supervised fine-tuning setting, the full ranking model outperforms the sliding window model, demonstrating the advantage of full ranking with proper fine-tuning.

% \MXY{Issues in full ranking}
% label 不完备；当输出更多的label时，这时候正负例label的不平衡问题显现出来，通常1:20的比例？
Furthermore, we identify the limitations of applying existing listwise training methods~\cite{rankvicuna, rankzephyr} to fine-tune a full ranking model: (1) Existing methods for ranking a passage list are primarily based on sliding windows, passing through the list from the back to the front. Theoretically, this approach can only guarantee the order of the top-ranked passages, rather than generating a full ranking list. (2) In terms of optimization objectives, existing methods use standard language modeling loss which applies equal penalty to all passage IDs in the training label. However, the training label for a full ranking model contains hundreds of passage IDs, with only a few of the top-ranked passage IDs being relevant. This leads to a significant imbalance between relevant and irrelevant passages. As a result, the penalty on relevant passage IDs is overshadowed by that on irrelevant ones, which conflicts with the evaluation of the ranking task that prioritizes top-ranked passages.

% In the supervised fine-tuning settings, existing methods~\cite{rankvicuna, rankzephyr} typically use the standard language modeling loss, which applies equal penalty to the passage IDs in the label. However, we deem that this loss does not align well with the ranking task, where top-ranked passage IDs should receive more attention during loss calculation. Applying equal penalties to all passage IDs causes the top-ranked ones to be overshadowed by others. We believe this issue is particularly significant when training a full ranking reranker, whose training labels include hundreds of passage IDs. 

% label construction and importance-aware loss。不长的话，和实验效果放一块，长的话分开
To address these limitations, we first design a multi-pass sliding window approach, which iteratively generates a full ranking list as training label, overcoming the limitation of producing only the top-ranked passages in a single pass. Then, we propose an importance-aware loss that reweights the passage IDs in the label based on their rank position, ensuring that higher-ranked passage IDs receive more attention. Extensive experiments on TREC and BEIR benchmarks demonstrate our fine-tuned full ranking model outperforms previous state-of-the-art models as well as our fine-tuned sliding window model. Figure~\ref{fig:intro_example} shows the advantage of our fine-tuned full ranking model compared to the sliding window model: an absolute improvement of 2.2 in terms of NDCG@10 and a reduction in latency by 29.3\% on TREC DL19 dataset.

% We conduct experiments based on state-of-the-art long-context LLMs, including open-source and proprietary models. 
% Our experiments are conducted on two well-established IR benchmarks: TREC~\cite{dl19} and BEIR~\cite{beir}. From the experimental results, we have the following observations: 
% Even though the full ranking model underperforms the sliding window model in the zero-shot setting, it consistently surpasses the sliding window model in the supervised fine-tuning scenario. 
% Figure~\ref{fig:intro_example} shows the advantage of our fine-tuned full ranking model compared to the sliding window model: an absolute improvement of 2.2 in terms of NDCG@10 and reduction in latency by 29.3\%.

Our contributions are summarized as follows:

\noindent$\bullet$~We conduct a comprehensive study of long-context LLMs for ranking. To the best of our knowledge, we are the first to investigate the application of long-context LLMs in ranking tasks.

\noindent$\bullet$~We reveal that, in a zero-shot setting, full ranking with long-context models is more efficient but less effective compared to the sliding window strategy.

\noindent$\bullet$~To enhance the effectiveness of full ranking, we propose a complete listwise label construction approach and a novel importance-aware learning objective for fine-tuning the full ranking model.

\begin{figure*}[!tb]
  \centering
  \vspace{0.3cm} % 增加图形上方的空白以往下移图形
  \includegraphics[width=.95\linewidth]{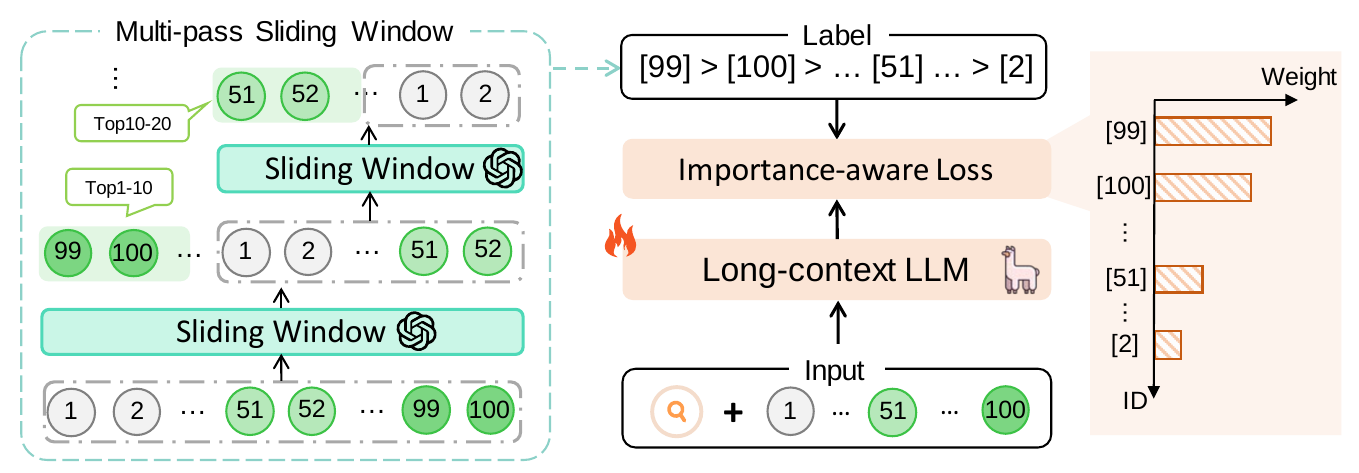}
  \caption{The training method of the full ranking model. We first use a multi-pass sliding window approach to iteratively obtain the full ranking list of passages. Then, we design an importance-aware loss that assigns different weights to the IDs in the label for model optimization.}
  \label{fig:model}
  % \vspace{-2.5cm}
\end{figure*}

\section{Listwise Ranking with LLMs}
In this section, we discuss the fundamentals of listwise ranking in zero-shot and supervised fine-tuning settings, respectively. 

\subsection{Zero-shot Ranking}
Given a user query $q$ and a list of passages $P=\left[p_1, \dots, p_N\right]$, the listwise ranking task takes $q$ and $P$ as the input, and outputs a ranked sequence of IDs indicating the relevance of each passage (\ie, [3] > [1] > \dots). The prompt we use is shown in Appendix~\ref{app:listwise_prompt}. In this paper, we use two different ranking strategies based on long-context LLMs: (1) the full ranking strategy and (2) the sliding window strategy. The full ranking involves inputting the entire passage list $P$ into the LLM, enabling it to rank all passages simultaneously in a single step. The sliding window strategy uses a window of size $w$, sliding from the end of the passage list to the beginning with a step size $s$. In this paper, following previous studies~\cite{rankgpt, rankvicuna}, we set passage number $N$, window size $w$, and step size $s$ to 100, 20, and 10, respectively.

\subsection{Supervised Fine-tuning} \label{sec:sft}
Existing studies~\cite{rankvicuna, rankzephyr} on fine-tuning listwise rerankers are primarily based on distillation techniques: the teacher model (\eg, GPT-4) receives a list of passages and outputs a sequence of passage IDs, which will be used as the supervised label for fine-tuning. The training process then optimizes the listwise reranker by minimizing the standard language modeling loss $\mathcal{L}$:
\begin{equation}
\label{equ:loss1}
\mathcal{L} = -\sum_{i=1}^{|y|} \log(P_{\theta}(y_i \mid x, y_{<i})),
\end{equation}
where $x$ represents the input prompt and $y$ is the teacher label, which is primarily obtained by re-ranking the top-20 passages retrieved by BM25.

\section{Fine-tuning Full Ranking Model} \label{sec:method}
As we mentioned in Section~\ref{sec:intro}, directly applying existing listwise training methods to train a full ranking model has two limitations: (1) One-pass sliding window process can only guarantee the top ranking, failing to obtain the full ranking list, similar to how a single pass of bubble sort algorithm can only guarantee the top-1 ranked item. Directly applying a full ranking to all passages is reasonable, but it is less effective than the sliding window strategy in zero-shot setting, which will be discussed in Section~\ref{subsec:effectiveness}. (2) Given a full ranking list, which exhibits a significant imbalance between relevant and irrelevant passage IDs, the standard language modeling loss applies equal penalties to each ID, causing the top-ranked IDs to be overwhelmed. To address these limitations, we design a multi-pass sliding window approach to generate high-quality full ranking lists as training labels and propose a novel importance-aware learning objective for model optimization. Next, we will introduce the details of our methods.

\subsection{Multi-pass Sliding Window Approach}
To generate the full ranking list for model fine-tuning, we first employ BM25 to retrieve the top-100 candidate passages, which are then reranked using a teacher model. In theory, a single pass of the sliding window process (ranking from back to front with a window size of 20 and a stride of 10) can only ensure the retrieval of the top-10 most relevant passages. To overcome this limitation, we propose a multi-pass sliding window approach to obtain the full ranking list, as illustrated in Figure~\ref{fig:model}.
In the first pass, the sliding window strategy is applied to rerank all 100 passages, yielding the top-10 most relevant passages. In the second pass, the same strategy is used to rerank the remaining 90 passages, producing the next 10 most relevant passages. This iterative process continues until the entire reranked list is constructed.
% Note that we choose to not use the teacher model to directly generate full ranking labels. The reason is that the full reranking task is much more challenging even for the teacher model, resulting in labels that are less effective compared to those obtained through a sliding window-based strategy.

% for two primary reasons: (1) The full reranking task is much challenging even for the teacher model, resulting in labels that are less effective compared to those obtained through a sliding window strategy. (2) Utilizing the same strategy for generating labels ensures a fair comparison between student models.

\subsection{Importance-Aware Learning Objective}
The full ranking label contains up to 100 passage IDs, with a very small proportion of relevant passage IDs. Using the standard language modeling loss leads to the loss contributions of top-ranked relevant passage IDs being overshadowed by others. This misaligns with the evaluation of ranking task, where more relevant passages are of greater importance and have a larger impact on ranking metrics. For example, ``[99]'' should be assigned the highest importance than other ids in the label ``[99] > [100] > $\cdots$'' while calculating the loss. To address this issue, we propose an importance-aware loss function $\mathcal{L}_{\text{ia}}$ which includes a position-based weight $w_p$ to reweight the importance of each passage ID in the label. The $\mathcal{L}_{\text{ia}}$ is defined as:
\begin{align}
\label{equ:loss2}
\mathcal{L}_{\text{ia}} &= -\sum_{i=1}^{|y|} w_i \log(P_{\theta}(y_i \mid x, y_{<i})), \\
w_i &=
\begin{cases} 
1 + \frac{1}{\log_2(p_i + 1)}, & i \in \text{passage IDs}, \\
\alpha, & i \notin \text{passage IDs}.
\end{cases}
\end{align}
Here, $p_i$ represents the passage rank corresponding to the $i$-th token. Note that each passage ID will be split into multiple tokens (\eg, [99] will be split into ``['', ``9'', ``9'', and ``]''), and these tokens have the same weight.\footnote{We also explore adding the passage IDs as new tokens into the LLM's vocabulary. Unfortunately, this strategy does not bring performance improvements.} Besides, we believe that the importance of passage IDs is higher than that of relational operator ``>'', so we set the weight of ``>'' as $\alpha$ ($\alpha <= 1$). By incorporating the importance-aware loss $\mathcal{L}_{\text{ia}}$, we ensure that higher-ranked passage IDs receive greater weight during loss calculation, which better aligns with the training of the full-ranking model.

% In standard language modeling, the loss for each passage ID is computed with equal weight. However, full ranking label contain 

% As the number of input passages increases and extends to the full length of the passage list, the loss contributions of top-ranked passage IDs can be overshadowed by those ranked lower, which misaligns with the essence of ranking tasks, where higher-ranked passage IDs should be assigned greater importance. 

% \paragraph{Training Label for Sliding Window}
% We use a powerful teacher model to obtain training labels for the sliding window model and full ranking model based on the same ranking window size 20 for fair comparison. As for the sliding window model (the model can only rank 20 passages at a time), following previous studies~\cite{rankvicuna, rankzephyr}, we use BM25 to retrieve top-20 passages and rerank them using the teacher model. The teacher-generated orderings are used as the training labels. We also experimented with sampling 20 passages from the top-100 retrieved ones as training passages, but found the performance inferior to directly using the top-20 passages. Detailed experimental results are presented in Table~\ref{tab:sample_strategy}. 

\section{Experiments}
\subsection{Setting}
\paragraph{Evaluation Datasets} \label{data}
For evaluation, we employed datasets from TREC DL 2019~\cite{dl19} and TREC DL 2020~\cite{dl20}, as well as BEIR~\cite{beir} benchmark. BEIR includes 18 datasets from various domains, designed to assess the zero-shot ability of ranking models. Following previous studies~\cite{rankgpt}, we choose 8 BEIR datasets for model evaluation. We rerank the top-100 passages retrieved by BM25 and use NDCG@10 as the evaluation metric.

\paragraph{Baselines} In addition to comparing the performance of sliding window model and full ranking model, we also include several finetuned rerankers for comparison: monoBERT (340M)~\cite{monobert}, monoT5 (220M)~\cite{monot5}, RankVicuna~\cite{rankvicuna} and RankZephyr~\cite{rankzephyr}. MonoBERT and monoT5 are trained on annotated query-passage pairs from the MS MARCO training set, while RankVicuna and RankZephyr are distilled from ranking lists generated by proprietary models, namely ChatGPT and GPT-4. The detailed descriptions of the baselines are presented in Appendix~\ref{app:baselines}.

\paragraph{Implementation Details}
In our experiments, we evaluate the performance of full ranking strategy based on long-context LLMs under two different settings: zero-shot and supervised fine-tuning, and compare with the sliding window strategy. 

For the zero-shot setting, we select both open-source and proprietary models to provide a comprehensive evaluation. The open-source models include Mistral-7B-Instruct-v0.3 (32k), LLaMA3.1-8B-Instruct (128k), and Qwen2.5-7B-Instruct (128k). These models have been chosen for their ability to handle long contexts and strong performance on a wide range of tasks. We also include two proprietary models, GPT-4o-mini and GPT-4o (both have 128k context)\footnote{The version of GPT-4o-mini and GPT-4o we used is gpt-4o-mini-2024-07-18 and gpt-4o-2024-08-06, respectively.}. 

For the supervised fine-tuning setting, we fine-tune a sliding window model for comparison with the full ranking model. Following previous studies~\cite {rankvicuna, rankzephyr}, we use BM25 to retrieve top-20 passages and rerank them using a teacher model to obtain the training label. We also experimented with sampling 20 passages from the top-100 retrieved ones as training passages but found the performance inferior to directly using the top-20 passages. Detailed experimental results are presented in Table~\ref{tab:sample_strategy}. We choose Mistral-7B-Instruct-v0.3 as our backbone model and use two different teacher models: GPT-4o-mini and GPT-4o. The finetuned full ranking model and sliding window model are denoted as $\textbf{RankMistral}_\textbf{100}$ and $\textbf{RankMistral}_\textbf{20}$, respectively. Both models are optimized with our proposed importance-aware loss for fair comparison. The detailed training configuration and API Cost for generating training labels can be found in appendix~\ref{app:train_config} and~\ref{app:api_cost}, respectively.

\begin{table*}[]
\centering
\small
\setlength{\tabcolsep}{0.8mm}{
\begin{tabular}{llccccccccccc}
\toprule
Models & Strategy & DL19 & \multicolumn{1}{c|}{DL20} & Covid & DBPedia & SciFact & NFCorpus & Signal & Robust04 & Touche & News & Avg. \\ \midrule
BM25 & - & 50.58 & \multicolumn{1}{c|}{47.96} & 59.47 & 31.80 & 67.89 & 33.75 & 33.04 & 40.70 & 44.22 & 39.52 & 43.80 \\
monoBERT (340M) & Pointwise & 70.72 & \multicolumn{1}{c|}{67.28} & 73.45 & 41.69 & 62.22 & 34.92 & 30.63 & 44.21 & 30.26 & 47.03 & 45.55 \\
monoT5 (220M) & Pointwise & 70.58 & \multicolumn{1}{c|}{67.33} & 75.94 & 42.43 & 65.07 & 35.42 & 31.20 & 44.15 & 30.35 & 46.98 & 46.44 \\
RankVicuna (7B) & Sliding & 67.72 & \multicolumn{1}{c|}{65.98} & 79.19 & 44.51 & 70.67 & 34.51 & 34.24 & 48.33 & 33.00 & 47.15 & 48.95 \\
RankZepyer (7B) & Sliding & 73.39 & \multicolumn{1}{c|}{70.02} & 82.92 & 44.42 & 75.42 & 38.26 & 31.41 & 53.73 & 30.22 & 52.80 & \textbf{51.15} \\ \midrule
\multicolumn{13}{c}{Zero-shot} \\ \midrule
\multirow{2}{*}{Mistral-v0.3 (7B)} & Sliding & 62.55 & \multicolumn{1}{c|}{58.48} & 74.79 & 40.66 & 56.61 & 34.68 & 30.09 & 45.61 & 34.41 & 44.40 & 45.16 \\
 & Full & 45.33 & \multicolumn{1}{c|}{47.86} & 62.63 & 33.85 & 51.58 & 32.71 & 27.63 & 37.38 & 37.54 & 37.76 & 40.14 \\
\multirow{2}{*}{Llama-3.1 (8B)} & Sliding & 65.18 & \multicolumn{1}{c|}{59.49} & 74.68 & 39.23 & 56.41 & 35.33 & 28.07 & 45.86 & 36.78 & 44.98 & 45.17 \\
 & Full & 53.50 & \multicolumn{1}{c|}{52.17} & 60.90 & 34.78 & 61.01 & 33.20 & 31.37 & 39.42 & \textbf{41.83} & 41.61 & 43.02 \\
\multirow{2}{*}{Qwen-2.5 (7B)} & Sliding & 68.34 & \multicolumn{1}{c|}{64.89} & 79.76 & 40.87 & 71.32 & 37.97 & 30.66 & 52.87 & 32.02 & 49.49 & 49.37 \\
 & Full & 57.44 & \multicolumn{1}{c|}{54.42} & 65.06 & 35.57 & 61.49 & 33.75 & 27.35 & 37.60 & 26.91 & 37.55 & 40.66 \\
\multirow{2}{*}{GPT-4o-mini} & Sliding & 72.36 & \multicolumn{1}{c|}{67.30} & 80.03 & 44.54 & 73.14 & 38.73 & 33.64 & 57.41 & 30.91 & 50.91 & 51.16 \\
 & Full & 68.80 & \multicolumn{1}{c|}{63.02} & 77.70 & 41.97 & 71.42 & 37.35 & 32.35 & 52.32 & 31.37 & 47.33 & 48.98 \\
\multirow{2}{*}{GPT-4o} & Sliding & \textbf{74.78} & \multicolumn{1}{c|}{69.52} & \textbf{83.41} & \textbf{45.56} & \textbf{77.41} & \textbf{39.67} & \textbf{34.20} & \textbf{60.25} & 32.26 & \textbf{51.92} & \textbf{53.09} \\
 & Full & 73.94 & \multicolumn{1}{c|}{\textbf{70.03}} & 82.10 & 43.31 & 74.85 & 39.00 & 32.63 & 55.95 & 30.42 & 47.96 & 50.78 \\ \midrule
\multicolumn{13}{c}{SFT from GPT-4o-mini} \\ \midrule
$\text{RankMistral}_{20}$ & Sliding & 69.08 & \multicolumn{1}{c|}{66.31} & 80.37 & 43.46 & 72.08 & 37.46 & 32.97 & 54.70 & 33.74 & 48.84 & 50.45 \\
$\text{RankMistral}_{100}$ & Full & \textbf{73.17} & \multicolumn{1}{c|}{\textbf{70.16}} & \textbf{82.57} & \textbf{44.54} & \textbf{75.47} & \textbf{38.73} & \textbf{33.49} & \textbf{56.36} & \textbf{38.77} & \textbf{51.08} & \textbf{52.63} \\ \midrule
\multicolumn{13}{c}{SFT from GPT-4o} \\ \midrule
$\text{RankMistral}_{20}$ & Sliding & 70.34 & \multicolumn{1}{c|}{69.58} & 80.86 & 42.52 & 75.82 & 38.38 & 33.90 & 54.69 & \textbf{37.18} & \textbf{51.42} & 51.85 \\
$\text{RankMistral}_{100}$ & Full & \textbf{72.55} & \multicolumn{1}{c|}{\textbf{71.29}} & \textbf{82.24} & \textbf{43.54} & \textbf{77.04} & \textbf{39.14} & \textbf{33.99} & \textbf{57.91} & 34.63 & 50.59 & \textbf{52.40} \\ \midrule
\multicolumn{13}{c}{w/o $L_{ia}$ (SFT from GPT-4o)} \\ \midrule
$\text{RankMistral}_{20}$ & Sliding & 71.14 & \multicolumn{1}{c|}{68.49} & 80.8 & 42.53 & \textbf{76.15} & 38.54 & 33.12 & 55.63 & 31.62 & \textbf{50.86} & 51.16 \\
$\text{RankMistral}_{100}$ & Full & \textbf{73.19} & \multicolumn{1}{c|}{\textbf{71.19}} & \textbf{81.24} & \textbf{42.71} & 75.82 & \textbf{38.85} & \textbf{34.0} & \textbf{57.40} & \textbf{33.21} & 50.27 & \textbf{51.69} \\ \bottomrule
\end{tabular}}
\caption{Results (NDCG@10) on TREC and BEIR. The best results among each part are marked in bold respectively. Avg. represents the averaged result of the 8 BEIR datasets.}
\label{tab:main_exp}
% \vspace{1mm}
\end{table*}

\subsection{Effectiveness Analysis} \label{subsec:effectiveness}
In this section, we evaluate the performance of full ranking strategy and sliding window strategy based on several long-context LLMs in zero-shot and fine-tuning settings. The results are shown in Figure~\ref{tab:main_exp}. For the three open-source long-context LLMs, we use abbreviations: Mistral-v0.3 (Mistral-7B-Instruct-v0.3), LLaMA-3.1 (LLaMA-3.1-8B-Instruct), and Qwen-2.5 (Qwen2.5-7B-Instruct). ``Sliding'' and ``Full'' refer to the sliding window and full ranking strategies, respectively. Next, we will provide a detailed analysis of both settings.

\paragraph{Zero-shot}
In the zero-shot setting, the full ranking strategy underperforms the sliding window strategy on nearly all datasets across all long-context LLMs. This demonstrates that full ranking significantly increases the ranking difficulty of LLM, resulting in a performance drop. However, there are some cases where the full ranking strategy outperforms the sliding window strategy. For instance, on the Touche dataset, the full ranking strategy yields better results for Mistral-v0.3, LLaMA-3.1, and GPT-4o-mini. This may be due to the dataset characteristics, where global interactions among passages play a larger role in assessing the relevance.

Besides, among the evaluated models, GPT-4o stands out by achieving the best full ranking performance. Notably, it achieves an average NDCG@10 of 71.99 on TREC and 50.78 on BEIR, proving its strong capability for processing long-context input in ranking tasks.

\begin{figure*}[!t]
        \vspace{-1mm}
        \centering
	\includegraphics[width=1\linewidth]{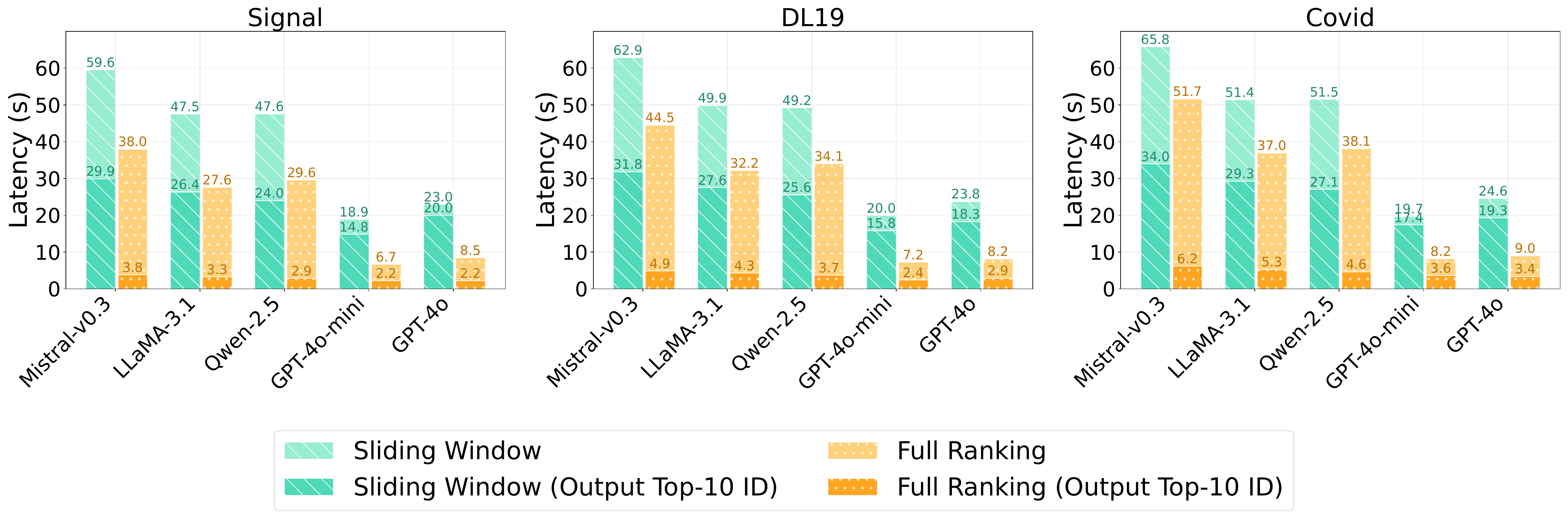}
	\caption{Latency of ranking top-100 passages based on full ranking and sliding window strategy. ``Output Top-10 ID'' indicates that the LLM only generates the top-10 ranked passage IDs.}
	\label{fig:latency}
     % \vspace{2mm}
\end{figure*}

\paragraph{Supervised Fine-tuning}
In supervised fine-tuning setting, we find that our full ranking model $\text{RankMistral}_{100}$ outperforms sliding window model $\text{RankMistral}_{20}$ on nearly all datasets. Notably, $\text{RankMistral}_{100}$ achieves an average improvement of about 4 and 2 points on TREC and BEIR, respectively, compared to $\text{RankMistral}_{20}$ when fine-tuned from GPT-4o-mini. These results indicate that task-specific fine-tuning can help the full ranking model better model the global interaction between all the passages, thus yielding better ranking performance than the sliding window model. Additionally, we also experiment with using Qwen2.5-7B-Instruct as the backbone model and observe the same conclusion. Detailed results are shown in Table~\ref{tab:qwen_exp}.

Furthermore, we conduct an ablation study on $\text{RankMistral}_{20}$ and $\text{RankMistral}_{100}$ to prove the effectiveness of our importance-aware loss $L_{ia}$. Both models are fine-tuned from GPT-4o and trained with standard language modeling loss. The results are shown in part ``w/o $L_{ia}$ (SFT from GPT-4o)''. After removing $L_{ia}$, both $\text{RankMistral}_{20}$ and $\text{RankMistral}_{100}$ drop about 0.7 points on BEIR Avg, proving that $L_{ia}$ makes the model focus more on top-ranked passage IDs, thereby enhancing its ranking effectiveness. Besides, even trained with standard language modeling loss, $\text{RankMistral}_{100}$ still outperforms $\text{RankMistral}_{20}$ on dl19, dl20, and BEIR Avg, which further demonstrates the effectiveness advantage of $\text{RankMistral}_{100}$ in supervised fine-tuning setting. 

Lastly, we also investigate the performance of $\text{RankMistral}_{100}$ under different ranking settings (\ie, different initial passage order and ranking numbers). Due to limited space, we present the detailed analysis in Appendix~\ref{app:passage_order}.

\subsection{Efficiency Analysis}
As discussed in Section~\ref{sec:intro}, the full ranking strategy eliminates the redundant and time-consuming sliding window operations. However, the significantly increased input length also introduces additional computational latency. In this part, we conduct experiments to measure the latency of two strategies.

Specifically, we rerank the top-100 passages retrieved by BM25 based on the sliding window strategy and full ranking strategy on three datasets—Signal, DL19, and Covid—which cover different passage lengths. In addition, as current search engines primarily display the top-10 search results, it is unnecessary to generate the whole list of passage IDs for real-time application. Thus, we also evaluate the latency of outputting only the top-10 passage IDs based on both strategies, which can be implemented by setting the model’s maximum number of output tokens. Note that in the sliding window process (with a window size of 20 and a step size of 10), only outputting 10 passage IDs at each step still ensures the final top-10 ranked IDs. 

We measure the latency on a 40GB Nvidia A100 GPU and average across all queries within each dataset. For GPT-4o-mini and GPT-4o, the latency is measured through API calls. Note that, since both $\text{RankMistral}{100}$ and $\text{RankMistral}{20}$ are fine-tuning on Mistral-7B-Instruct-v0.3, we do not compare their efficiency separately.

The results are presented in Figure~\ref{fig:latency}. The following observations can be drawn from the figure: 

(1) Ranking latency of full ranking strategy (light orange bar) is much smaller than sliding window strategy (light green bar) across all long-context LLMs. For example, on the Signal dataset, the full ranking strategy achieves an efficiency improvement of approximately 42\% on Qwen-2.5 and 65\% on GPT-4o compared to the sliding window strategy. This demonstrates that full ranking strategy is more efficient than sliding window strategy. 

(2) When the models are restricted to output only the top-10 passage IDs, both strategies (dark green bar and dark orange bar) demonstrate significant latency reduction, indicating that the number of decoded tokens greatly impacts the overall latency. Furthermore, the latency gap between the two strategies becomes even more pronounced. For example, on the Signal dataset, the sliding window strategy takes 29.9 seconds, while the full ranking strategy only takes 3.8 seconds, resulting in an approximate 8x speed-up. This highlights the efficiency advantage of the full ranking strategy in real-world search engines, where only the top-10 results need to be displayed after a user submits a query. Besides, the lower latency of GPT-4o and GPT-4o-mini compared to the other open-source models may be due to differences in inference resources, such as hardware acceleration and model optimization.

\subsection{Impact of Passage Number $N$}
\begin{figure}[t]
    \vspace{-2mm}
    \centering
    \begin{subfigure}[b]{1\linewidth}
        \centering
        \includegraphics[width=\linewidth]{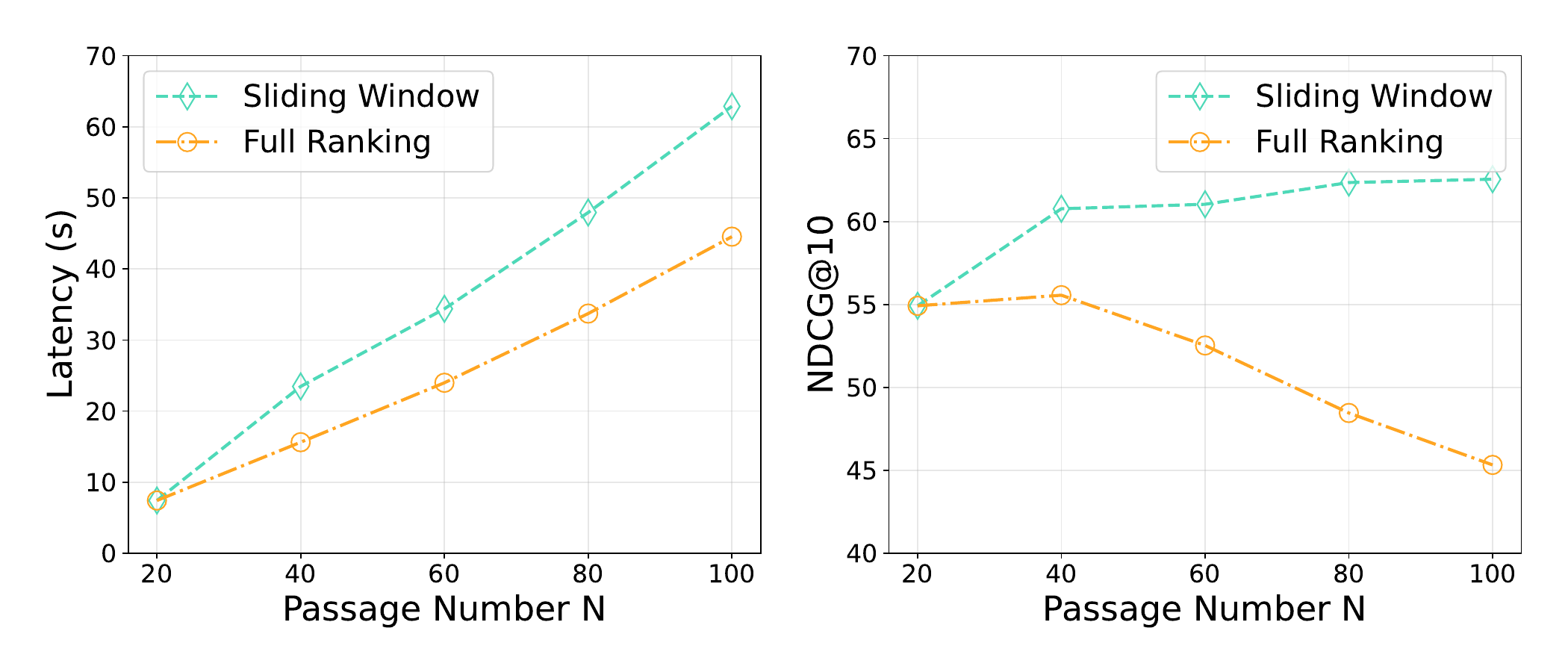}
        \caption{Results of Mistral-7B-instruct-v0.3.}
    \end{subfigure}
    
    % \vspace{1em} % 添加垂直间距

    \begin{subfigure}[b]{1\linewidth}
        \centering
        \includegraphics[width=\linewidth]{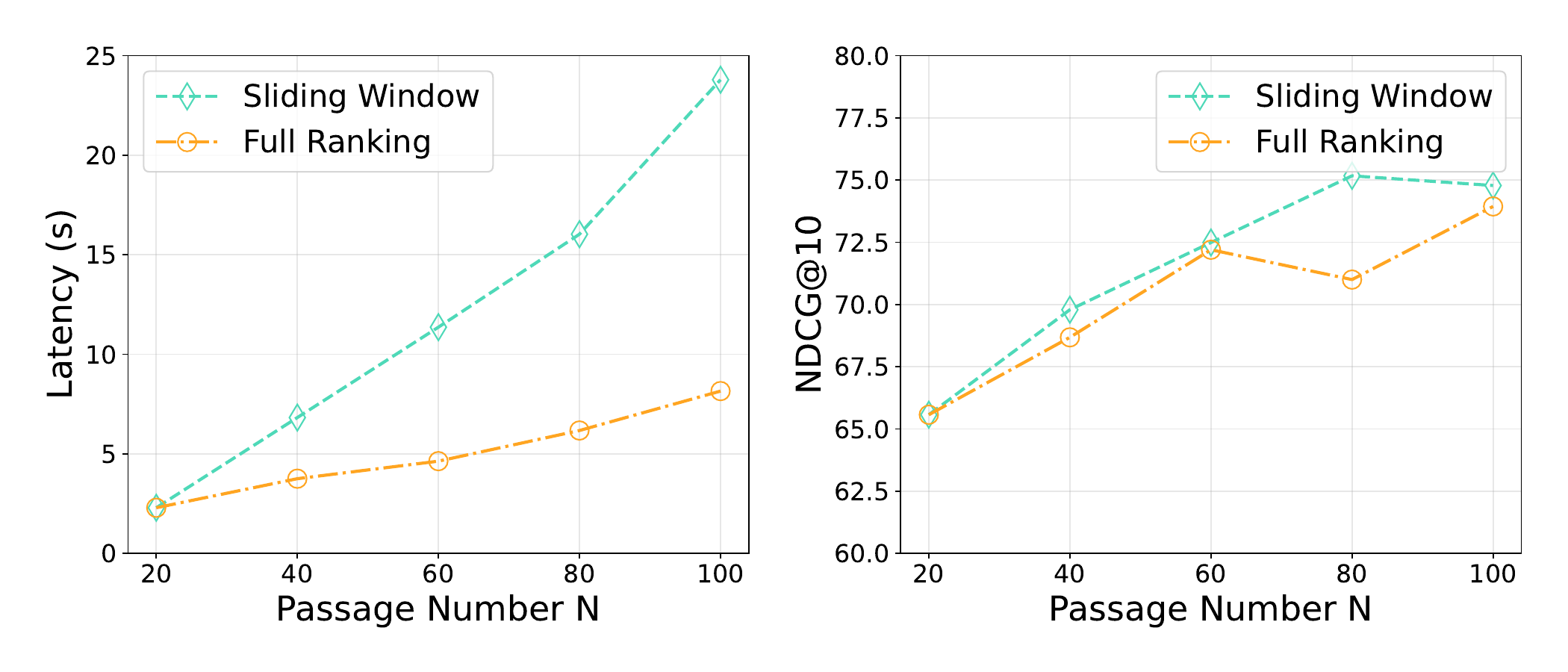}
        \caption{Results of GPT-4o.}
    \end{subfigure}

    \caption{Comparison of sliding window strategy and full ranking strategy on DL19 dataset based on Mistral-7B-instruct-v0.3 and GPT-4o, respectively.}
    \label{fig:different_N}
    \vspace{2mm}
    \end{figure}

\paragraph{Efficiency and Effectiveness Analysis}
In this part, we further compare the efficiency and effectiveness of full ranking strategy and sliding window strategy across different numbers of passages $N$. We choose different values of $N \in \{20, 40, 60, 80, 100\}$ and conduct the experiments on DL19 dataset based on open-source model Mistral-7B-instruct-v0.3 and the proprietary model GPT-4o respectively. The results are shown in Figure~\ref{fig:different_N}. From the results, we have the following observations: (1) For different $N$, the latency of the full ranking strategy consistently remains lower than that of the sliding window strategy, with the gap becoming more noticeable as $N$ increases. (2) As $N$ increases, the latency of full ranking exhibits an almost linear growth, which benefits from the optimization within LLMs for handling long contexts. (3) Across various values of $N$, the sliding window strategy demonstrates better effectiveness compared to the full ranking strategy, similar to the phenomenon observed in Table~\ref{tab:main_exp}. Additionally, due to GPT-4o’s strong ability to model long contexts, the effectiveness gap between the two strategies is relatively smaller compared to Mistral-7B-Instruct-v0.3.

% \usepackage{multirow}
% \usepackage{multirow}

\paragraph{Generalization of $\text{RankMistral}_\text{100}$}
As we mentioned in Section~\ref{sec:method}, our full ranking model is trained to rank 100 passages at a time. However, it remains unclear whether it can generalize to other numbers of passages ($N$). In this section, we selected different values of $N \in \{20, 40, 60, 80, 100\}$ and tested the performance of $\text{RankMistral}_\text{100}$ based on full ranking strategy, comparing it with $\text{RankMistral}_\text{20}$ using sliding window strategy. Both models are fine-tuned using our importance-aware loss $L_{ia}$ with labels generated by GPT-4o. We conducted experiments on DL19, DL20, and BEIR, and show the results in Table~\ref{tab:mistral_generalization}. From the results, it can be seen that $\text{RankMistral}_{100}$ performs better than $\text{RankMistral}_{20}$ across different $N$ values. This indicates that even though $\text{RankMistral}_{100}$ is trained on passage lists of length 100, it can still generalize to variable-length ranking tasks. The comprehensive results are presented in Table~\ref{tab:mistral_generalization_all}.

\begin{figure}[!t]
    \vspace{-2mm}
	\centering
	\includegraphics[width=1\linewidth]{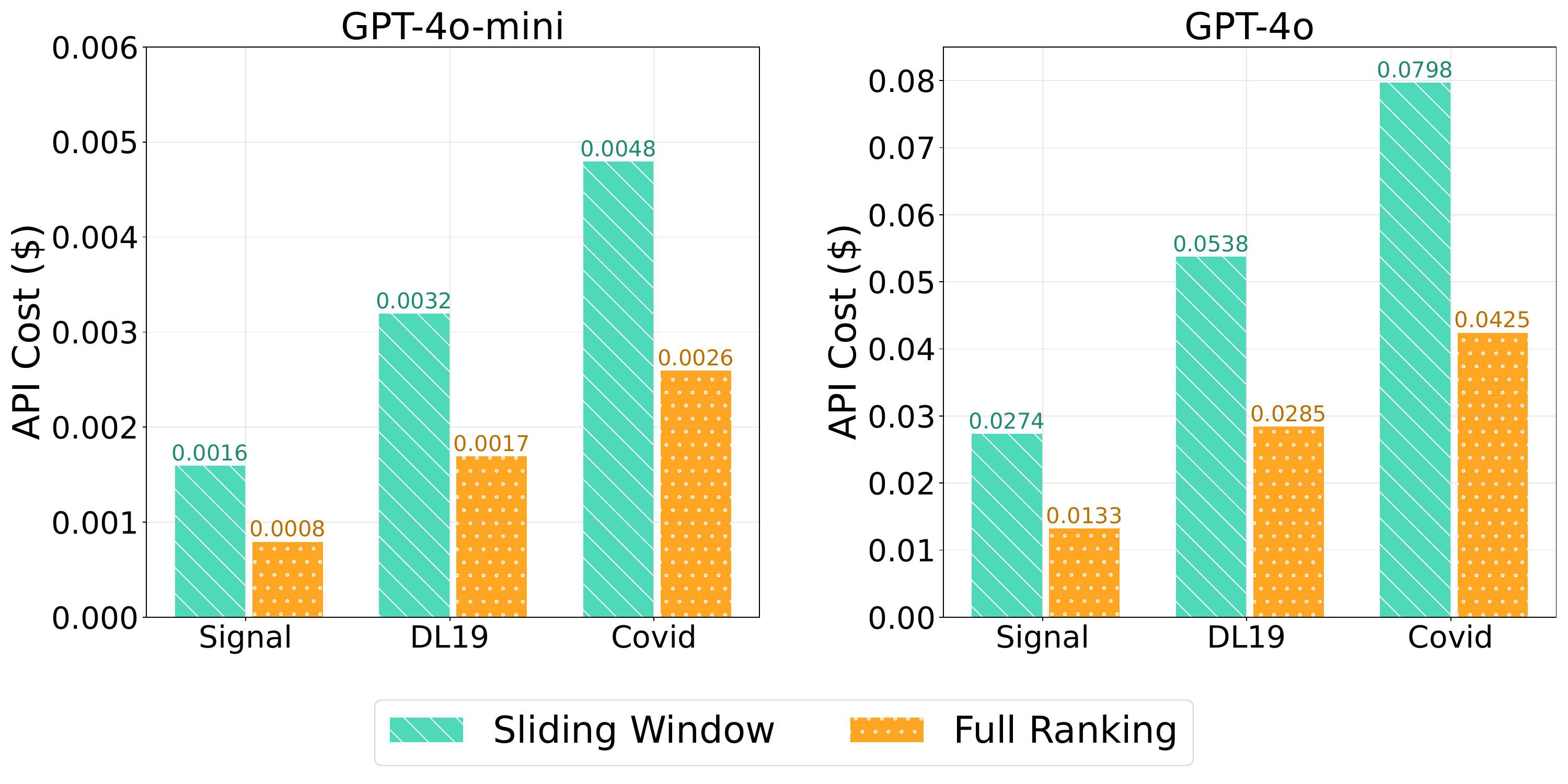}
	\caption{The comparison of API cost per query of sliding windows strategy and full ranking strategy when ranking top-100 retrieved passages.}
	\label{fig:api_cost}
    \vspace{1mm}
\end{figure}

\begin{table}[]
\small
\centering
\setlength{\tabcolsep}{2.5mm}{
\begin{tabular}{llccc}
\toprule
N & \multicolumn{1}{c}{Model} & DL19 & DL20 & BEIR Avg \\ \midrule
\multirow{2}{*}{20} & $\text{RankMistral}_{20}$ & 63.47 & 59.56 & 48.45 \\
 & $\text{RankMistral}_{100}$ & \textbf{65.82} & \textbf{62.03} & \textbf{49.87} \\ \midrule
\multirow{2}{*}{40} & $\text{RankMistral}_{20}$ & 65.89 & 63.88 & 50.07 \\
 & $\text{RankMistral}_{100}$ & \textbf{67.77} & \textbf{66.01} & \textbf{51.88} \\ \midrule
\multirow{2}{*}{60} & $\text{RankMistral}_{20}$ & 70.34 & 64.82 & 50.66 \\
 & $\text{RankMistral}_{100}$ & \textbf{70.97} & \textbf{68.74} & \textbf{52.00} \\ \midrule
\multirow{2}{*}{80} & $\text{RankMistral}_{20}$ & 70.28 & 68.29 & 51.13 \\
 & $\text{RankMistral}_{100}$ & \textbf{71.36} & \textbf{70.22} & \textbf{52.14} \\ \midrule
\multirow{2}{*}{100} & $\text{RankMistral}_{20}$ & 70.34 & 69.58 & 51.85 \\
 & $\text{RankMistral}_{100}$ & \textbf{72.55} & \textbf{71.29} & \textbf{52.40} \\ \bottomrule
\end{tabular}}
\caption{Results (NDCG@10) of $\text{RankMistral}_\text{100}$ and $\text{RankMistral}_\text{20}$ under different passage number $N$.}
\label{tab:mistral_generalization}
\vspace{1mm}
\end{table}

\subsection{API Cost Comparison}
As we mentioned in Section~\ref{sec:intro}, full ranking avoids the inference of redundant passages between adjacent windows in sliding window strategy, thereby reducing the API cost for LLMs. In this section, we investigate the API cost savings of the full ranking strategy compared to the sliding window strategy. We selected the Signal, DL19, and Covid datasets, and used GPT-4o-mini and GPT-4o as the inference models. We rerank the BM25-retrieved top-100 passages and the API costs are calculated based on the number of input and output tokens, using OpenAI’s official price. As shown in Figure~\ref{fig:api_cost}, full ranking strategy reduces API costs by about 50\% compared to the sliding window strategy, which proves its cost-efficiency. The differences in API costs across datasets are mainly due to varying passage lengths, with longer passages leading to higher costs.

\section{Related Work}
\subsection{LLMs for Passage Ranking}
The application of LLMs into information retrieval~\cite{llm4ir_survey} has sparked significant interest, leading to numerous studies on utilizing LLMs for passage ranking tasks. Existing approaches leveraging LLMs for passage ranking can be classified into three categories based on their ranking strategies: pointwise, pairwise, and listwise methods. Pointwise methods~\cite{liang2022holistic, sachan2022improving, beyond_yes_no, demorank} assess the relevance between a query and each passage independently. Pairwise methods~\cite{qin2023large, PRP-Graph} involves comparing two passages at a time to determine which one is more relevant to the query. Listwise methods~\cite{rankgpt, rankvicuna, first, pe_rank} directly rank a list of passages. By taking multiple passages as input, listwise methods perform multi-passage comparisons and show promising ranking performance. Given the promising performance of listwise ranking, there has been a growing interest in exploring this approach. For example, ~\citet{rankgpt} introduce a prompt-based framework that utilizes ChatGPT to rank passages in a zero-shot manner. ~\citet{rankvicuna} and ~\citet{rankzephyr} further propose to distill the strong ranking ability of ChatGPT or GPT-4 into moderate-size LLMs. To accelerate the listwise ranking process, ~\citet{pe_rank} propose to replace the input passage context with single passage embedding, thereby reducing the total input length.

% ~\citet{first} propose to use the logits of the first generated passage ID to obtain the ranked list of the passages. To accelerate the listwise ranking process, ~\citet{pe_rank} propose to replace the input passage context with single passage embedding, thereby reducing the total input length.

Due to limited input length of LLMs, existing listwise methods mainly take a subset of passages as input and apply sliding window strategy to rank passages from back to front. While this approach yields promising performance, it still suffers from efficiency and cost problems as mentioned in Section~\ref{sec:intro}. Long-context LLMs, with the ability to process significantly longer inputs, present new opportunities for listwise passage ranking. However, the academic community has yet to explore the application of long-context LLMs in passage ranking. In this paper, we make the first attempt to investigate this direction.

\subsection{Long Context Large Language Models}
Recently, there has been significant progress in developing LLMs that can process longer input sequences. The context window sizes of LLMs have grown from 1024 tokens in GPT-2~\cite{gpt2} to 4096 in LLaMA 2~\cite{llama2}. To address the computational challenges of longer contexts, methods like efficient attention mechanisms~\cite{LongNet, Beacon, landmark} and positional interpolation~\cite{Positional_Interpolation} have been proposed. For example, Landmark attention~\cite{landmark} extends LLaMA 7B's context length from 4K to 32K by using "landmark tokens" to represent context blocks and fine-tuning attention to select relevant tokens. Additionally, ~\citet{parallel_window} propose chunking long contexts into sub-windows and reusing positional embeddings, enabling the model to handle longer contexts without extra fine-tuning. Building upon these techniques, many long-context LLMs have been developed and released, such as Mistral-7B-instruct-v0.3 (32k)~\cite{mistral}, Qwen2.5-7B-Instruct (128k)~\cite{qwen2} and GPT-4o (128k). Although some previous studies~\cite{Retrieval_meets_lllm, LongBench} have discussed the impact of long-context LLMs on retrieval, less attention has been paid to ranking tasks. In this work, we intend to provide a comprehensive discussion on the potential benefits and challenges of long-context LLMs for ranking tasks.

\section{Conclusion}
In this paper, we conduct a comprehensive study on passage ranking with long-context LLMs. Our experiments demonstrate that, the full ranking strategy is more efficient than sliding window strategy. After fine-tuning, full ranking model also outperforms sliding window model in ranking effectiveness. For fine-tuning full ranking model, we propose a multi-pass sliding window approach and a importance-aware learning objective for label generation and model optimization. Experiments demonstrate the effectiveness of our method.

\section*{Limitations} \label{limitation}
We acknowledge some limitations in our work. Firstly, due to our limited resources, we cannot experiment with larger open-source long-context LLMs, such as those with 30B or even 70B parameters. Investigating the impact of model size on the effectiveness and efficiency of full ranking strategy would be an interesting direction for future research. Secondly, the efficiency advantage of the full ranking strategy is attributed to the existing long-context LLMs. We did not design a specialized long-context LLM for ranking task. However, we consider this a promising direction and will treat it as future work.

% \bibliographystyle{acl_natbib}
% \bibliography{custom}

\clearpage
\appendix

\section{Listwise Ranking Prompt} \label{app:listwise_prompt}
The listwise ranking prompt used in this work is shown as bellow:
\begin{tcolorbox}[colback=gray!10, colframe=black, title=Prompt: Listwise Ranking Prompt]
You are RankLLM, an intelligent assistant that can rank passages based on their relevancy to the query.

I will provide you with \texttt{\{num\}} passages, each indicated by a numerical identifier \texttt{[]}.

Rank the passages based on their relevance to the search query: \texttt{\{query\}}.

\texttt{[1]} \{passage 1\}  

\texttt{[2]} \{passage 2\}

...

\texttt{[\{num\}]} \{passage \{num\}\} \\

Search Query: \texttt{\{query\}}. \\

Rank the \texttt{\{num\}} passages above based on their relevance to the search query. All the passages should be included and listed using identifiers, in descending order of relevance. The output format should be \texttt{[]} > \texttt{[]}, e.g., \texttt{[4]} > \texttt{[2]}. Only respond with the ranking results, do not say any word or explain.
\end{tcolorbox}

\section{Training Configuration} \label{app:train_config}
\begin{figure}[!t]
	\centering
	\includegraphics[width=0.9\linewidth]{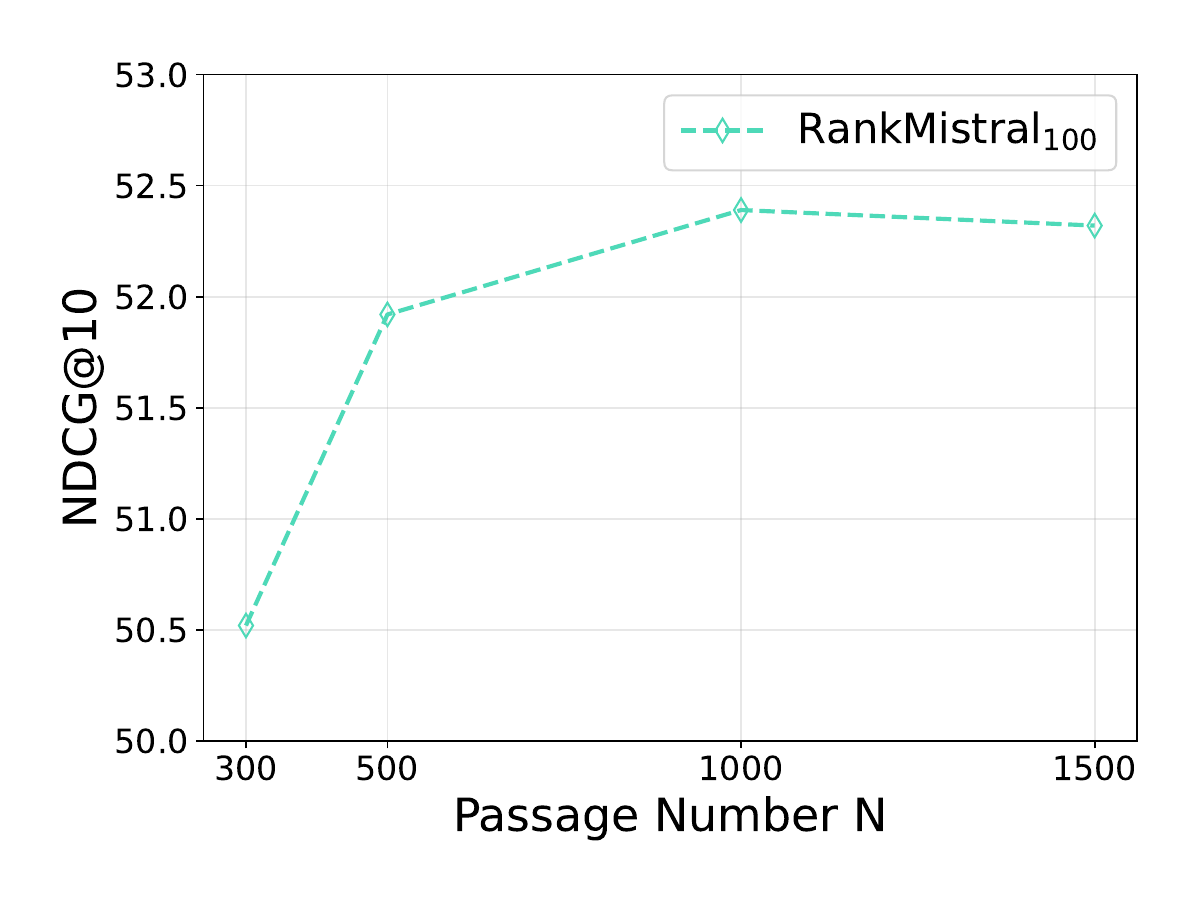}
	\caption{The results (NDCG@10) of $\text{RankMistral}_{100}$ on BEIR Avg using different numbers training queries. The figure shows that increasing the number of training queries from
1000 to 1500 does not improve performance on BEIR Avg.}
	\label{fig:training_size}
\end{figure}

Our backbone model Mistral-7B-Instruct-v0.3 for supervised-fine-tuning is available at \url{https://huggingface.co/mistralai/Mistral-7B-Instruct-v0.3}. We apply the prompt shown in Figure~\ref{app:listwise_prompt} to construct the training input. We randomly sample 1k queries from MS MARCO training set to generate the teacher labels. During our experiments, we find that using more queries for training does not yield better results (see Figure~\ref{fig:training_size}). Both $\text{RankMistral}_\text{20}$ and $\text{RankMistral}_\text{100}$ are finetuned for 4 epochs with a learning rate of 5e-6 and batch size of 1. The hyper-parameter $\alpha$ used in Equation~\ref{equ:loss2} is set as 1. Following RankZephyr~\citet{rankzephyr}, we apply noisy embeddings~\cite{NEFTune} and bfloat16 precision during model training. We conduct all the experiments on 4 A100-40G GPUs. 

\begin{table}[]
\small
\centering
\setlength{\tabcolsep}{1.2mm}{
\begin{tabular}{lccc}
\toprule
Method & NDCG@1 & NDCG@5 & NDCG@10 \\ \midrule
BM25 & 54.26 & 52.78 & 50.58 \\
$\text{RankMistral}_{100}$ & 72.87 & 74.19 & 72.55 \\ \midrule
\multicolumn{4}{l}{Initial passage order} \\
Random order & 20.93 & 19.26 & 20.34 \\
Reverse order & 3.88 & 4.69 & 5.34 \\ \midrule
Number of ranking &  &  &  \\
Rank 2 times & 74.42 & 76.46 & 73.92 \\
Rank 3 times & 74.42 & 77.53 & 74.30 \\
Rank 4 times & \textbf{74.42} & \textbf{77.61} & \textbf{74.50} \\ \bottomrule
\end{tabular}}
\caption{The impact of initial passage order and number of ranking on $\text{RankMistral}_{100}$ on TREC DL19.}
\label{tab:passage_order}
\end{table}

\section{The Impact of Passage Order and Ranking Number} \label{app:passage_order}
Previous studies have revealed that the initial passage order and ranking number have an impact on the ranking performance. In this section, we explore the ranking performance of our trained $\text{RankMistral}_{100}$ on the TREC DL19 dataset under different initial order and ranking number settings to gain further insights into full ranking.

Regarding the initial passage order, we experimented with two different configurations: random order and reverse order. The results in Figure~\ref{tab:passage_order} show that the initial order of passages significantly impacts the ranking performance, suggesting that a good initial order is important for listwise ranking. This is consistent with the conclusions from previous work~\cite{rankgpt, tourrank}. There are also some training techniques aimed at enhancing the robustness of listwise rankers to the initial order, such as shuffling the order of training passages during training; however, this is not the focus of our paper. We also observe that increasing the ranking number improves the ranking performance. However, when the ranking number reaches 3 or 4, the performance gains start to converge.

\begin{table*}[]
\centering
\small
\setlength{\tabcolsep}{5mm}{
\begin{tabular}{lcc}
\toprule
Model & \$USD / 1K Input Tokens & \$USD / 1K Output Tokens \\ \midrule
GPT-4o-mini-2024-07-18 & 0.00015 & 0.00060 \\
GPT-4o-2024-08-06 & 0.0025 & 0.0100 \\ \bottomrule
\end{tabular}}
\caption{The price of GPT-4o-mini-2024-07-18 and GPT-4o-2024-08-06.}
\label{tab:api_price}
\end{table*}

\begin{table*}[]
\centering
\small
\setlength{\tabcolsep}{1.3mm}{
\begin{tabular}{llcc|ccccccccc}
\toprule
N & Model & DL19 & DL20 & Covid & DBPedia & SciFact & NFCorpus & Signal & Robust04 & Touche & News & Avg. \\ \midrule
\multirow{2}{*}{20} & $\text{RankMistral}_{20}$ & 63.47 & 59.56 & 69.98 & 37.78 & 74.42 & 36.46 & 32.29 & 48.49 & 40.55 & \textbf{47.61} & 48.45 \\
 & $\text{RankMistral}_{100}$ & \textbf{65.82} & \textbf{62.03} & \textbf{70.86} & \textbf{38.73} & \textbf{75.64} & \textbf{37.11} & \textbf{35.33} & \textbf{50.67} & \textbf{43.43} & 47.17 & \textbf{49.87} \\ \midrule
\multirow{2}{*}{40} & $\text{RankMistral}_{20}$ & 65.89 & 63.88 & 75.91 & 40.75 & 75.40 & 36.91 & 33.35 & 52.10 & 39.00 & 47.06 & 50.07 \\
 & $\text{RankMistral}_{100}$ & \textbf{67.77} & \textbf{66.01} & \textbf{77.46} & \textbf{41.79} & \textbf{76.79} & \textbf{38.25} & \textbf{35.13} & \textbf{53.27} & \textbf{43.05} & \textbf{49.31} & \textbf{51.88} \\ \midrule
\multirow{2}{*}{60} & $\text{RankMistral}_{20}$ & 70.34 & 64.82 & 78.04 & 41.43 & 74.79 & 37.97 & 33.38 & 54.31 & 37.02 & 48.30 & 50.66 \\
 & $\text{RankMistral}_{100}$ & \textbf{70.97} & \textbf{68.74} & \textbf{79.96} & \textbf{41.92} & \textbf{76.70} & \textbf{38.66} & \textbf{34.18} & \textbf{55.93} & \textbf{38.45} & \textbf{50.18} & \textbf{52.00} \\ \midrule
\multirow{2}{*}{80} & $\text{RankMistral}_{20}$ & 70.28 & 68.29 & 80.15 & 42.04 & 75.91 & 37.86 & \textbf{33.77} & 54.65 & 35.09 & 49.53 & 51.13 \\
 & $\text{RankMistral}_{100}$ & \textbf{71.36} & \textbf{70.22} & \textbf{81.25} & \textbf{42.38} & \textbf{76.35} & \textbf{38.95} & 33.47 & \textbf{56.67} & \textbf{36.96} & \textbf{51.10} & \textbf{52.14} \\ \midrule
\multirow{2}{*}{100} & $\text{RankMistral}_{20}$ & 70.34 & 69.58 & 80.86 & 42.52 & 75.82 & 38.38 & 33.90 & 54.69 & \textbf{37.18} & \textbf{51.42} & 51.85 \\
 & $\text{RankMistral}_{100}$ & \textbf{72.55} & \textbf{71.29} & \textbf{82.24} & \textbf{43.54} & \textbf{77.04} & \textbf{39.14} & \textbf{33.99} & \textbf{57.91} & 34.63 & 50.59 & \textbf{52.40} \\ \bottomrule
\end{tabular}}
\caption{Results (NDCG@10) on TREC and BEIR under different passage number $N$. Best performing models are marked in bold. All the models are finetuned with labels generated by GPT-4o.}
\label{tab:mistral_generalization_all}
\end{table*}

\section{API Cost} \label{app:api_cost}
In Table~\ref{tab:api_price}, we present the price for two different teacher models we used in this paper. We use 1k query to generate the teacher label for different models. For fine-tuning the sliding window model $\text{RankMistral}_{20}$, we re-rank the top-20 passages retrieved by BM25 using the teacher model, costing about \$1.7 for GPT-4o-mini and \$29 for GPT-4o. For the full ranking model, we use the multi-pass sliding window strategy to obtain the full ranked list, with costs of around \$15.7 for GPT-4o-mini and \$261 for GPT-4o.

\section{Baselines Details} \label{app:baselines}
The baselines we used for comparison are:

\noindent$\bullet$ \textbf{monoBERT}~\cite{monobert}: A BERT-large based cross-encoder re-ranker, fine-tuned using the MS MARCO dataset~\cite{MSMARCO}.

\noindent$\bullet$ \textbf{monoT5}~\cite{monot5}: A re-ranker using a sequence-to-sequence model with T5 to determine relevance scores.

\noindent$\bullet$ \textbf{RankVicuna}~\cite{rankvicuna}: A listwise reranker fine-tuned from GPT-3.5 generated ranked list. 

\noindent$\bullet$ \textbf{RankZephyr}~\cite{rankzephyr}: A listwise reranker distilled from GPT-3.5 and GPT-4 with a two-stage training process.

\begin{table*}[]
\centering
\small
\setlength{\tabcolsep}{0.7mm}{
\begin{tabular}{lccc|ccccccccc}
\toprule
Teacher & Sampling Strategy & DL19 & DL20 & Covid & DBPedia & SciFact & NFCorpus & Signal & Robust04 & Touche & News & Avg. \\ \midrule
\multirow{2}{*}{GPT-4o-mini} & Top-20 & \textbf{69.20} & \textbf{68.66} & 79.72 & \textbf{43.54} & \textbf{73.93} & 37.49 & \textbf{32.74} & \textbf{54.95} & \textbf{32.69} & 50.21 & \textbf{50.66} \\
 & Sample 20 & 69.13 & 67.49 & \textbf{80.44} & 42.06 & 73.04 & \textbf{37.97} & 32.26 & 54.19 & 31.50 & \textbf{50.30} & 50.22 \\ \midrule
\multirow{2}{*}{GPT-4o} & Top-20 & \textbf{71.14} & \textbf{68.49} & 80.80 & \textbf{42.53} & \textbf{76.15} & \textbf{38.54} & \textbf{33.12} & \textbf{55.63} & \textbf{31.62} & \textbf{50.86} & \textbf{51.16} \\
 & Sample 20 & 69.55 & 65.78 & \textbf{81.71} & 41.21 & 75.35 & 37.71 & 31.70 & 55.57 & 30.71 & 48.44 & 50.30 \\ \bottomrule
\end{tabular}}
\caption{The performance (NDCG@10) of $\text{RankMistral}_\text{20}$ on TREC and BEIR based on different sampling strategy. Top-20 refers to the top-20 passages retrieved by BM25, while sample 20 indicates a random sampling of 20 passages from the top-100 passages retrieved by BM25.}
\label{tab:sample_strategy}
\end{table*}

\begin{table*}[]
\centering
\small
\setlength{\tabcolsep}{1.1mm}{
\begin{tabular}{llcc|ccccccccc}
\toprule
Models & Strategy & DL19 & DL20 & Covid & DBPedia & SciFact & NFCorpus & Signal & Robust04 & Touche & News & Avg. \\ \midrule
$\text{RankQwen}_{20}$ & Sliding & \textbf{72.79} & 67.96 & 81.18 & 42.33 & 75.79 & \textbf{38.53} & 31.10 & \textbf{56.97} & 30.93 & 49.06 & 50.74 \\
$\text{RankQwen}_{100}$ & Full & 71.45 & \textbf{68.74} & \textbf{82.64} & \textbf{42.84} & \textbf{76.30} & 38.23 & \textbf{32.45} & 55.93 & \textbf{36.00} & \textbf{49.51} & \textbf{51.74} \\ \bottomrule
\end{tabular}}
\caption{The performance comparison between finetuned sliding window model and full ranking model (denoted as $\text{RankQwen}_{20}$ and $\text{RankQwen}_{100}$, respectively) when using Qwen2.5-7B-Instruct as backbone model. The training labels for both models are generated by GPT-4o. The results indicate that $\text{RankQwen}_{100}$ outperforms $\text{RankQwen}_{20}$ on most datasets and achieves an average of 1-point improvement on BEIR Avg.}
\label{tab:qwen_exp}
\end{table*}

\end{document}